\documentclass[%
 reprint,
 amsmath,amssymb,
 aps,
]{revtex4-2}

\usepackage{graphicx}
\usepackage{dcolumn}
\usepackage{bm}
\usepackage[rawfloats=true]{floatrow} 
\restylefloat{figure}     

\begin{document}

\preprint{APS/123-QED}

\title{Application of neural network for exchange-correlation functional interpolation}

\author{Alexander Ryabov}
 \email{ryabov.alexandr@phystech.edu}
 \affiliation{Center for Design, Manufacturing and Materials, Skolkovo Institute of Science and Technology, Bolshoy Boulevard 30, bld. 1, Moscow, 143026, Russia}
 \affiliation{Moscow Institute of Physics and Technology (State University), Institutskiy per. 9, Dolgoprudny, Moscow Region 141700, Russia}
\author{Petr Zhilyaev}%
 \email{p.zhilyaev@skoltech.ru}
\affiliation{%
 Center for Design, Manufacturing and Materials, Skolkovo Institute of Science and Technology, Bolshoy Boulevard 30, bld. 1, Moscow, 143026, Russia\\
}%

\date{\today}

\begin{abstract}
Density functional theory (DFT) is one of the primary approaches to get a solution to the many-body Schrodinger equation. The essential part of the DFT theory is the exchange-correlation (XC) functional, which can not be obtained in analytical form. Accordingly, the accuracy improvement of the DFT is mainly based on the development of XC functional approximations. Commonly, they are built upon analytic solutions in low- and high-density limits and result from quantum Monte Carlo or post-Hartree-Fock numerical calculations. However, there is no universal functional form to incorporate these data into XC functional. Various parameterizations use heuristic rules to build a specific XC functional. The neural network (NN) approach to interpolate the data from higher precision theories can give a unified path to parametrize an XC functional. Moreover, data from many existing quantum chemical databases could provide the XC functional with improved accuracy. In this work, we develop NN XC functional, which gives both exchange potential and exchange energy density. Proposed NN architecture consists of two parts NN-E and NN-V, which could be trained in separate ways, which adds additional flexibility. We also show the suitability of the developed NN XC functional in the self-consistent cycle when applied to atoms, molecules, and crystals.   
\end{abstract}

\maketitle


\section{Introduction}
Since its emergence, density functional theory (DFT)~\cite{hohenberg1964inhomogeneous, kohn1965self}
serves as one of the primary methods of solving the many-body Schrodinger equation. The main theoretical bottleneck in DFT theory is the unknown form of the exchange-correlation (XC) functional. Therefore, the progress in developing more accurate XC functionals reveals more possibilities for using DFT in cases where high accuracy of quantum-mechanical calculations is required. Information for constructing XC functionals is taken from numerical calculations using quantum Monte Carlo or post-Hartree-Fock~\cite{ceperley1980ground, zhao2005exchange}. The influential Monte Carlo (MC) simulations of the uniform electron gas (UEG) by Ceperley and Adler~\cite{ceperley1980ground} led to the creation a number of practical local density approximations (LDAs)~\cite{vosko1980accurate, perdew1981self, perdew1992accurate}. The next big success in reaching better accuracy was achieved  by the generalized gradient approximation (GGA), which takes into consideration local gradients of electron density~\cite{wang1991spin, perdew1992atoms, perdew1996generalized}. This improvement sufficiently increased the capability of DFT to characterize systems with inhomogeneous electron densities. Calculations made by post-Hartree-Fock methods were also used to improve the quality of the XC functionals for molecular systems~\cite{zhao2005exchange, mardirossian2017thirty}. The search for advanced XC functional is ongoing, and still a very active direction of research~\cite{lani2016adiabatic, maier2016new, mori2014derivative, mori2018exact}.

Regardless of the evident triumph of the LDA and GGA, their development is a highly complex procedure that involves many heuristics stages. The XC functional's form is generally specified by physical insights (the local nature of the interaction, perturbation approach, analytical solutions in extreme cases, etc.), and a set of adjustable parameters~\cite{perdew1992accurate, perdew1996generalized}. Such inflexible functional form makes it difficult to integrate more numerical results from modern quantum MC~\cite{needs2009continuum, kolorenvc2011applications} and  post-Hartree-Fock calculations~\cite{cremer2011moller}, which could lead to increased XC functional accuracy. Therefore it may be fruitful to use an adaptive XC functional form that, on the one hand, facilitates the incorporation of numerical calculations and, on the other hand, enable to include the physical insights into it.

\begin{figure*}
    \centering
    \includegraphics[scale=0.6]{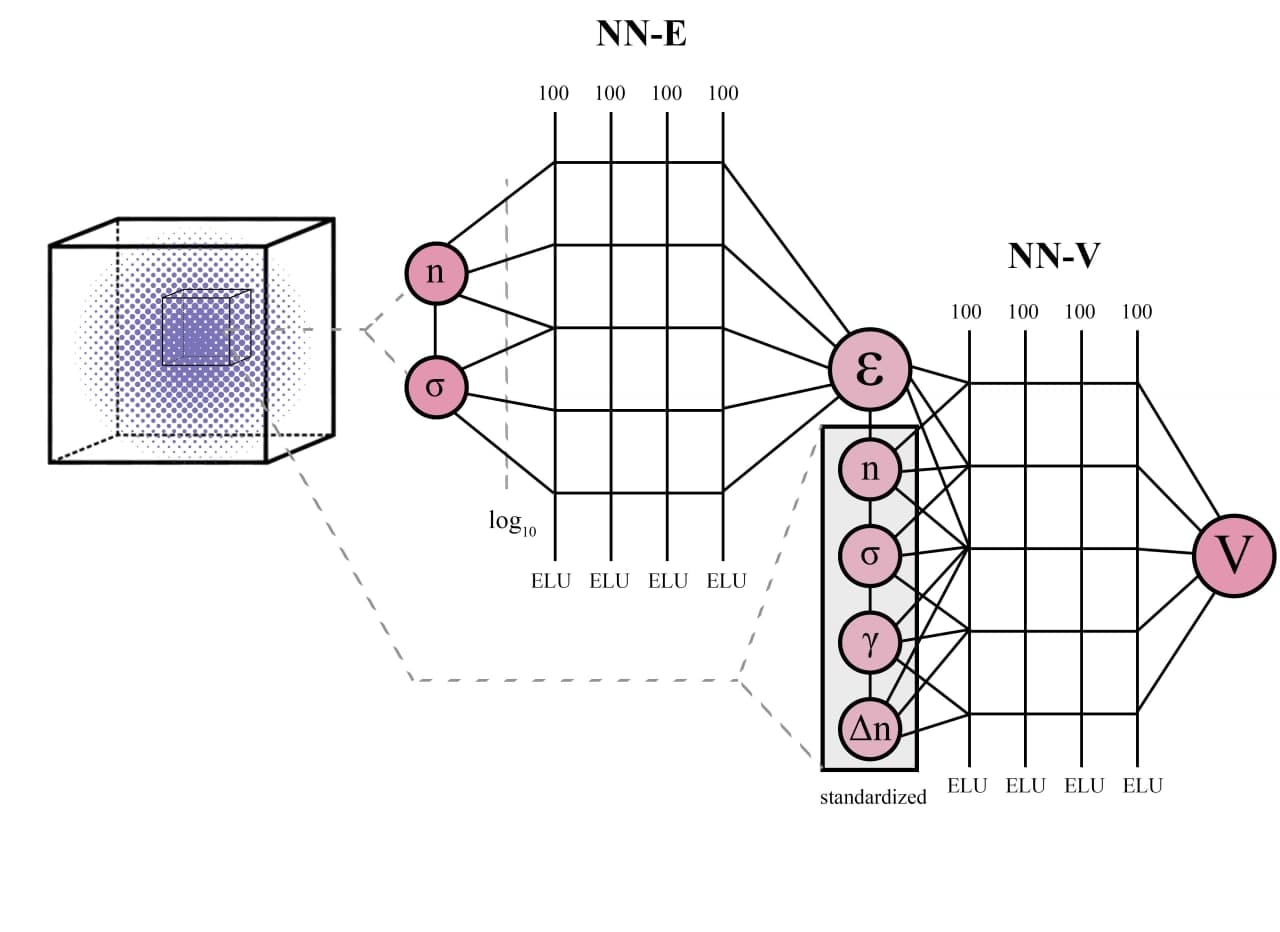}
    \caption{\textbf{Topology of the developed neural network.}}
    \label{fig:nnstructure}
\end{figure*}

One perspective candidate for the flexible XC functional form is the neural network (NN), which provides a universal approach to approximate any functional relationship~\cite{cybenko1989approximation}. Analytical information could be also included in the NN as a synthetically generated part of the data set. NN was first utilized as a functional form for XC potential by Tozer et al.~\cite{tozer1996exchange}. After that, several studies have been addressed the possibility of using NN to approximate XC functionals form~\cite{nagai2018neural, nagai2020completing, lei2019design,  ramos2019static, ryabov2020neural, li2021kohn}. Work by Nagai and co-authors~\cite{nagai2020completing} is especially worthy of note. They presented the first working NN XC functional, which gives better accuracy than traditional ones.

Despite significant advances in the development of XC functionals based on NNs, a wide range of issues remain. Namely, how to relate the exchange-correlation energy density ($\epsilon_{xc}$) and the exchange-correlation potential (XC potential, $V_{xc}$)? In the analytical description, such connection is provided by a standard differentiation, which is sometimes tedious but straightforward. But when new features are included in the NN, for example, the logarithm of density (standard feature scaling procedure), it is not clear how to incorporate them into the connection between XC potential and corresponding energy density. It is also essential to find a way to include physical insight into XC functionals, such as asymptotic analytical solutions and conservation laws. Another issue is an optimal NN architecture and feature selection for presenting XC functional. So far, no detailed comparison has been made of various NN architectures on the same training data.

We focus on constructing NN XC functionals that output both $\epsilon_{xc}$ and $V_{xc}$. The developed NN consists of two parts: one part is used to evaluate $\epsilon_{xc}$ (NN-E), and another approximates $V_{xc}$ (NN-V). They connected in such a way that the output of NN-E is one of the input features for NN-V (see fig. \ref{fig:nnstructure}). The train/test data set was obtained from the DFT calculation of crystalline silicon, benzene, and ammonia with PBE XC functional~\cite{perdew1996generalized, PhysRevLett.78.1396}. After XC NN training, we implemented it into existing DFT code~\cite{andrade2015real, andrade2012time, andrade2013real} and conducted self-consistent cycle calculations of train/test systems and in addition with atoms and molecules from IP13/03. The mean relative error of total and XC energies on training/test samples is on the order of $0.001\%$. The same errors on crystals and molecules that were not used to train XC NN increased by the order magnitude but were still small, around $0.01\%$. Such a small relative error indicates that the proposed architecture could be successfully used for XC functional form representation. The key feature of the proposed NN architecture is that the weights of NN-V are pre-trained on $\epsilon_{xc} \rightarrow V_{xc}$ mapping and fixed during learning on new training data. 

\begin{figure*}
    \centering
    \includegraphics[scale=0.6]{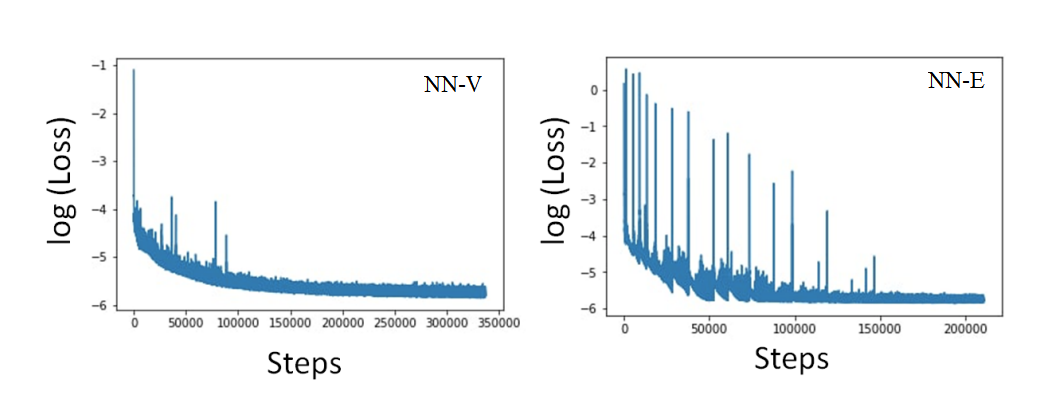}
    \caption{\textbf{Training curves of obtained neural network. The left one corresponds to NN-V training, where Log(Loss) is the logarithm of loss defined in "Methods" section. The right one corresponds to NN-E training with boundary conditions. }}
    \label{fig:training_curves}
\end{figure*}

It allows being sure that the output of NN-E is indeed the $\epsilon_{xc}$, because the relation between $\epsilon_{xc}$ and $V_{xc}$ is preserved. It is also should be noted that the boundary conditions are included by using extra datasets.They are synthetically generated to fulfill given boundary conditions. In our case it was just  $\epsilon_{xc} \rightarrow 0$ given that electron density ($n$) is vanishing, and $\epsilon_{xc} \rightarrow \epsilon_{xc}^{LDA}$ given that gradient modulus squared of electron density ($\sigma$) is also approaching zero.

This paper is outlined as follows: In methods section, we describe the data generation process, NN topology, features and  hyperparameters. In the results section, we first show the results of NN training, features distributions, and evaluated metrics; in the following, we demonstrate NN XC's application in DFT calculations on atoms and molecules that were not in the dataset. In conclusion section, we summarize our results and suggest following research steps that could be done with the developed NN XC pipeline, with a particular emphasis on integrating data from high-level methods such as quantum MC and post-HF calculation. 

\section{Methods}
The proposed neural network architecture consists of two parts: the NN-E and the NN-V. The NN-E serves to obtain the $\epsilon_{xc}$, and the NN-V allows calculating of the $V_{xc}$ from the corresponding energy using the output of NN-E. Only the spin  unpolarized case is considered. 

The NN-E input parameters are electron density $n$ and the square of the electron density gradient \eqref{sigma}. We use base-10 logarithmic transformation for preprocessing input features of NN-E. The input features for the NN-V are $\epsilon_{xc}$, $n$, \eqref{sigma}, \eqref{gamma} and Laplacian of the electron density $\Delta n$;

\begin{align}
  \label{sigma} \sigma &=\langle \nabla n, \nabla n \rangle,\\ 
  \label{gamma} \gamma &= \langle \nabla \sigma, \nabla n \rangle
\end{align}

\noindent
All NN-V features except $\epsilon_{xc}$ are converted to zero mean and unit variance (see. fig \ref{fig:nnstructure}).

At the first stage of training, only the NN-V is trained. In this case, the energy that is supplied to the input to the NN-V is obtained using the libxc ~\cite{lehtola2018recent} package. The first stage aims to teach mapping between the $\epsilon_{xc}$ and the corresponding potential, and the $\epsilon_{xc}$ known in advance is used. In this case the following loss function is applied:

\begin{align}
Loss = \frac{1}{N}\sum\limits_{i=1}^N(V_{xc}^{true}[i] - V_{xc}^{predicted}[i])^2
\end{align}

At the second stage, the NN-V weights are frozen, and only NN-E part is trained, but loss depends on the output of NN-V  ($V_{xc}$). The second stage aims to train the neural network to map electron density, $\sigma$ and $V_{xc}$. Simultaneously, the frozen NN-V provides the correct connection between $\epsilon_{xc}$ and $V_{xc}$. At this stage we additionally include boundary condition in the loss function:

\begin{align}
Loss = \frac{1}{N}\sum\limits_{i=1}^N[(V_{xc}^{true}[i] - V_{xc}^{predicted}[i])^2 + \nonumber\\ (\epsilon_{xc}^{predicted}(0, \sigma)[i] - 0)^2 +  \nonumber \\ (\epsilon_{xc}^{predicted}(n, 0)[i] - \epsilon_{xc}^{LDA}(n)[i])^2]
\label{bc_loss}
\end{align}

\begin{figure*}
    \centering
    \includegraphics[scale=0.56]{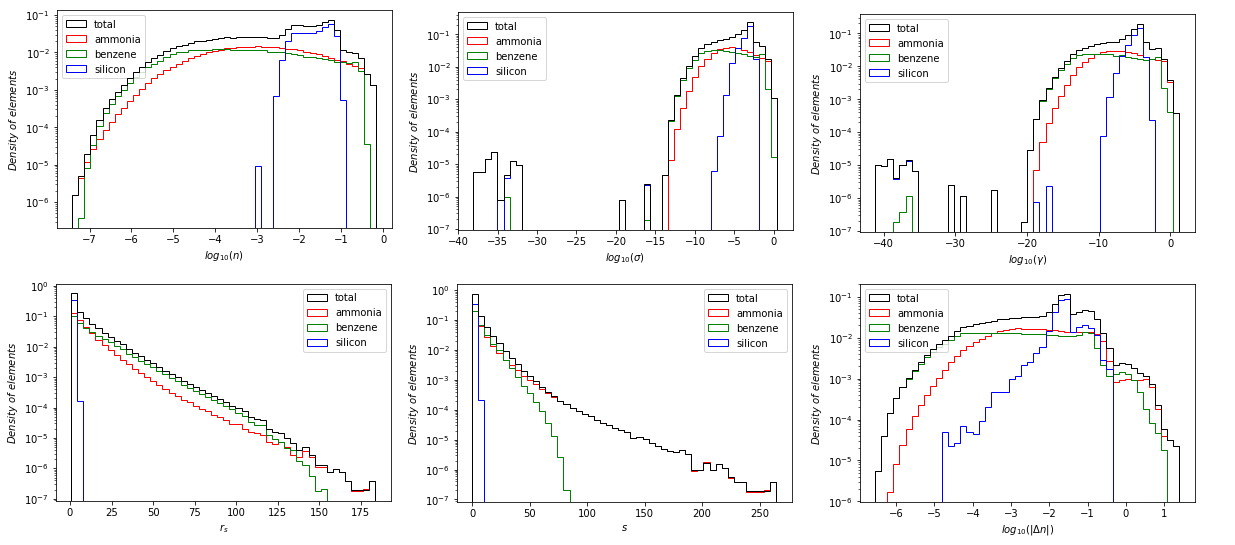}
    \caption{\textbf{Distributions of input features}}
    \label{fig:feature_distrib}
\end{figure*}

\begin{figure*}
    \centering
    \includegraphics[scale=0.3]{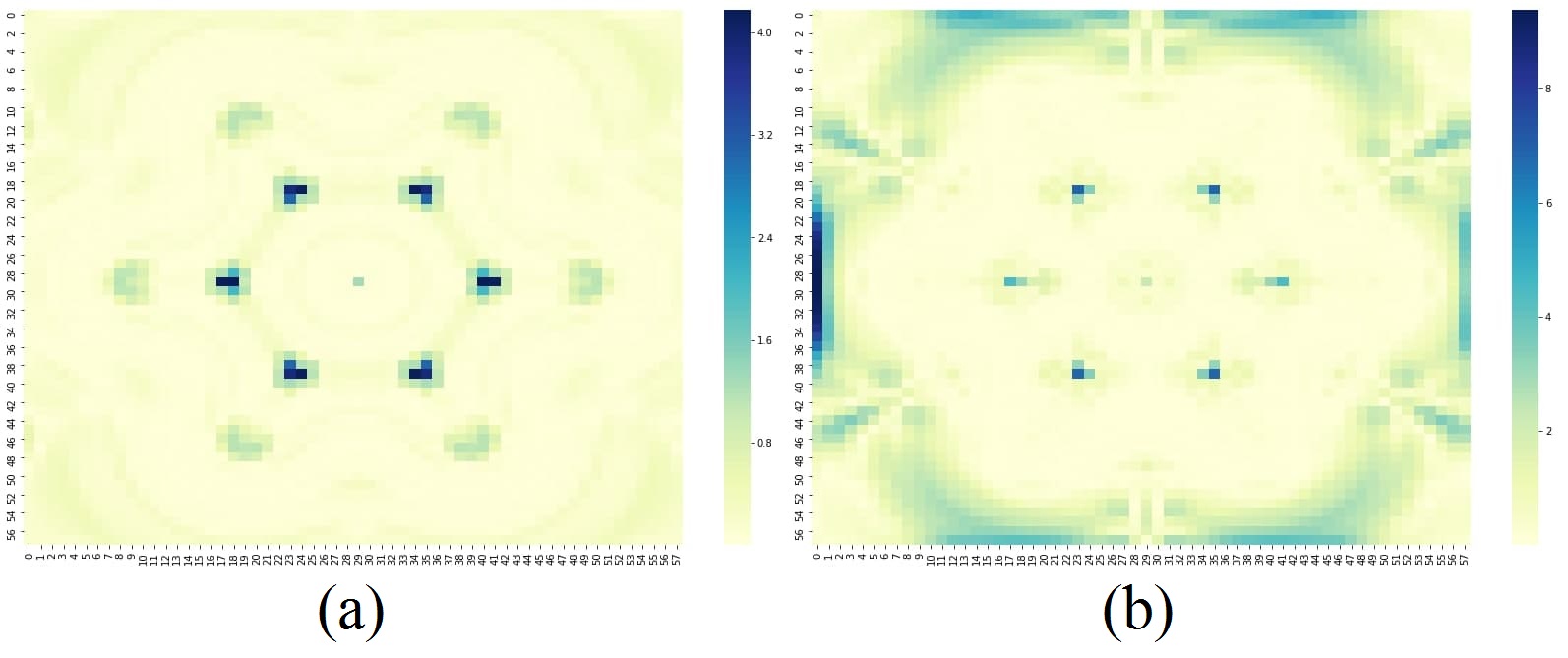}
    \caption{\textbf{Distributions of absolute error on the slice of benzene molecule}}
    \label{fig:benzene_distrib}
\end{figure*}

The first boundary condition is based on the fact that $\epsilon_{xc}$ tends to zero at zero density with any input $\sigma$. The second boundary condition follows from the fact that with zero $\sigma$ and any density  $\epsilon_{xc}$ leads to the corresponding energy for the case of the local density approximation.

\begin{figure*}
    \includegraphics[scale=0.7]{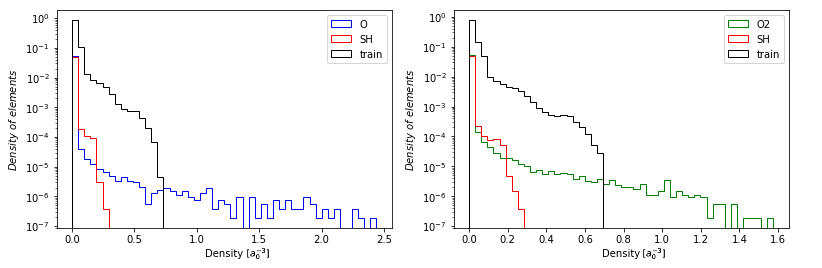}
    \caption{\textbf{Comparison of maximum electron density in SCF cycle between "good" case with small Exc error (SH) and "bad" case with large Exc error (O and O2) with respect to the training distribution }}
    \label{fig:O_SH_compare}
\end{figure*}

\begin{figure*}
    \includegraphics[scale=0.4]{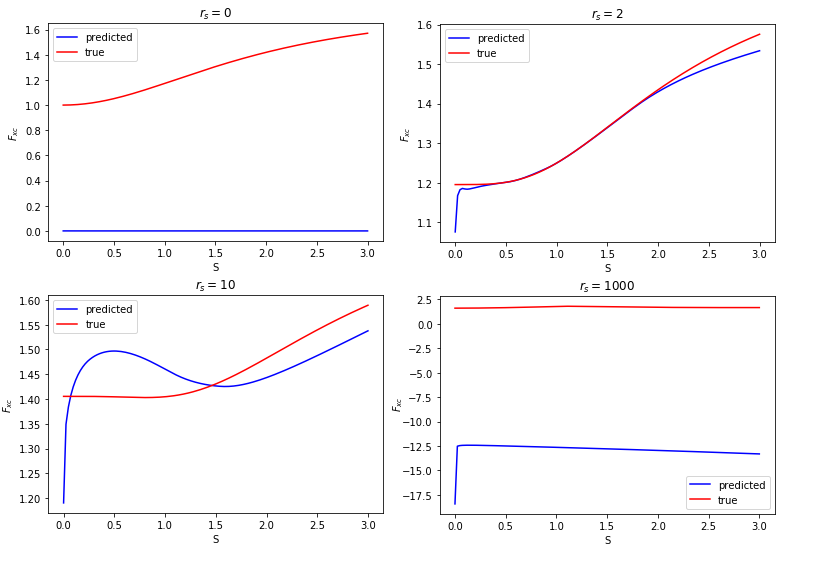}
    \caption{\textbf{Comparison of enhancement factor ($F_{xc} (r_s, s)$) obtained from NN-E (blue line) and libxc (red line) }}
    \label{fig:enhancement}
\end{figure*}


To obtain the training data, we carry out DFT calculations~\cite{hohenberg1964inhomogeneous, kohn1965self} of silicon, benzene, and ammonia. We perform the calculations in real space using the Octopus code~\cite{andrade2015real, andrade2012time, andrade2013real}. The XC functionals used include the exchange and correlation parts from~\cite{perdew1996generalized, PhysRevLett.78.1396}. The total number of the calculations is 10 for each chemical substance. Atomic configurations in these calculations are differed by applied strain $\pm 5 \%$. Grid spacing in the range from zero to 10 is used. This corresponds to $64 \times 64 \times 64$ mesh for silicon and benzene, and $65 \times 65 \times 65$ mesh for ammonia. The mesh size for ammonia is a little bit larger due to technical issues related to non-periodic boundary conditions. For avoiding gradient inconsistency in the boundaries we use cropping. We remove data points lying at a distance of 4 (due to numerical scheme of differentiation) or less from the boundaries of the parallelepiped. Finally, the dataset size was approximately 5.2 million samples.



The Pytorch framework~\cite{NEURIPS2019_9015} is utilized for training the neural network. We use the Adam algorithm for training with a learning rate descending from 0.001 by 25 percent every 20 epoch, and mean square error (MSE) loss. The batch size selected for training is 5000. The neural architecture used for the NN-E is a fully connected network with two input neurons, one linear output neuron, and four exponential linear units (ELU)~\cite{clevert2016fast} hidden layers with 100 neurons. NN-V has the same architecture except that the number of input neurons increased to 5. 

To implement boundary conditions \eqref{bc_loss} we also included two additional batches in each training step. The first one consists of zero electron density and non-zero gradients with corresponding zero $\epsilon_{xc}$. The second one contains non-zero electron density and zero gradients with corresponding $\epsilon_{xc}^{LDA}$.

\section{Results}

\subsection{Training of Neural Network}

Training curves for NN-V and NN-E are presented in Fig. \ref{fig:training_curves}. Jumps on the training curves are related with adjustable learning rate used for optimization algorithm. The loss reaches a plateau after 300000 batches for NN-V and 200000 batches for NN-E correspondingly. Each batch consisted of 32 sample. The final loss achieved values of about $10^-6$ in both cases.

The distributions of input features are presented in fig \ref{fig:feature_distrib}. In addition, the distributions of the Wigner-Seitz radius ($r_{s}=\left(\frac{3}{4 \pi n}\right)^{1 / 3}$, where $n$ is electron density) and the dimensionless gradient s ($s=\frac{\left(\frac{1}{n}\right)^{4 / 3}|\nabla n|}{2 \sqrt[3]{3} \pi^{2 / 3}}$) are presented. Individual distributions of substances included in the training dataset are normalized to the total number of elements in the entire training dataset. They are not uniform due to real systems used in training procedure. Analysis of such distributions is important to determine the limits of NN XC functional applicability, i.e. to detect cases where NN XC functional will certainly fail due to absence of such data in the training data set. This is especially important for case of heavy atoms in which core electron density could have large derivatives.

\begin{table}[b]
    \centering
    \begin{tabular}{lcccc}
        \multicolumn{1}{c}{\texttt{}}&
        \multicolumn{1}{c}{\texttt{$V_{xc}^{MSE}$}}& 
        \multicolumn{1}{c}{\texttt{$V_{xc}^{MAE}$}}&
        \multicolumn{1}{c}{\texttt{$\epsilon_{xc}^{MSE}$}}&
        \multicolumn{1}{c}{\texttt{$\epsilon_{xc}^{MAE}$}}\cr
        \hline
        Benzene& $1\cdot10^{-6}$ & $6\cdot10^{-4}$ & $4\cdot10^{-7}$ & $2\cdot10^{-4}$\\
        Silicon& $1\cdot10^{-6}$ & $4\cdot10^{-4}$ & $2\cdot10^{-5}$ & $1\cdot10^{-3}$\\
        Ammonia& $3\cdot10^{-6}$ & $1\cdot10^{-3}$ & $2\cdot10^{-6}$ &\ $4\cdot10^{-4}$
    \end{tabular}
    \caption{\textbf{Results of NN XC functional on a training dataset. }}
    \label{tab:train_set_results}
\end{table}

The training of NN was performed on benzene, silicon and ammonia. For each substance ten calculation was performed. In each calculation the system was stretched or compressed in range of $\pm{5 \%}$ of lattice constant to obtain electron densities in a wider range. 

\begin{table}
    \centering
    \begin{tabular}{lcc}
        \multicolumn{1}{c}{\texttt{}}&
        \multicolumn{1}{c}{\texttt{$E_{total}$ error, \%}}&
        \multicolumn{1}{c}{\texttt{$E_{xc}$ error, \%}}\cr
        \hline
        C& 0.052& 0.287\\
        S& 0.015& 0.102\\
        SH& 0.021& 0.007\\
        Cl& 0.010& 0.035\\
        OH& 0.554& 1.036\\
        $Cl_2$& 0.001& 0.035\\
        O& 1.556& 4.029\\
        P& 0.008& 0.071\\
        $O_2$& 0.646& 2.482\\
        PH& 0.014& 0.050\\
        $PH_2$& 0.011& 0.023\\
        $S_2$& 0.023& 0.040\\
        Si& 0.003& 0.253
    \end{tabular}
    \caption{\textbf{Results of testing NN XC functional on a subset of IP13/03 dataset. $E_{total}$ and $E_{xc}$ denote total energy and exchange-correlation energy correspondingly obtained after convergence. }}
    \label{tab:ip13_results}
\end{table}

The resulting MSE (mean square error) and MAE (mean average error) on training data set is presented in the table~\ref{tab:train_set_results}. Minimum MAE of $\epsilon_{xc}$ is obtained on benzene, maximum MAE is achieved on silicon. You can see in Figure \ref{fig:feature_distrib} that the distributions of silicon are significantly different from the overall distributions of the training dataset. We think that it is the explanation of such a high error on the training silicon subsample.

\subsection{Testing of Neural Network}

We selected IP13/03 dataset for testing proposed NN XC functional. All calculations are performed in $17.5 \times 17.5 \times 17.5~a_0$ parallelepiped with a $65 \times 65 \times 65$ mesh, which corresponds to a spacing of $0.273~a_0$.

We incorporate the developed NN XC functional into Octopus code and perform self-consistent cycle calculations. All calculations are converged, and relative total and xc energies errors are calculated. As a reference, we take PBE functional that was used for training of NN. Results are presented in the table \ref{tab:ip13_results}. One can see that the highest error is achieved for oxygen.

The reasons for the error on oxygen were analyzed. It turned out that the main reason for such a discrepancy on some substances, including oxygen, is the intermediate densities occurred in the process of a self-consistent cycle that the neural network did not see during the training process. The problem is that such densities can be arbitrarily large. The distributions of maximum electron densities in some cases are presented in Fig. \ref{fig:O_SH_compare} together with the training distribution. Analytical exchange-correlation functional contains asymptotics at infinity, which make it possible to correctly process such cases. However, during our training cycle, only boundary conditions at zero were used, but asymptotics at infinity were not used. This important fact describes obtained discrepancy and gives hope for the further development and improvement of neural network exchange-correlation functionals by including such asymptotics.

We also performed the in-plane analysis of absolute error of $\epsilon(xc)$ and $V_{XC}$ on a benzene molecule (see Fig. \ref{fig:benzene_distrib}). The maximum absolute errors are achieved in regions near carbon ions. We attribute this error to the high value of gradients and abrupt changes of electron density.

Enhancement factor $F_{xc}(r_s, s)$ obtained from NN-E prediction was tested. It compared with corresponding $F_{xc}$ obtained from prediction of analytical exchange-correlation functional (PBE) by libxc (See Fig. \ref{fig:enhancement}). There is a high similarity between the enhancement factors at densities that closely match the most frequent densities in the training set ($r_s$ = 2). We observe slightly less similarity at $r_s = 10$. This can be explained by the decrease in the number of elements with such a density in the training set.  In the limiting cases of high and low densities, a significant discrepancy is observed. In the case of high density (small $r_s$), this is explained by the absence of asymptotic at infinity. Despite the presence of asymptotic at zero densities in the process of training a neural network, the obtained result shows that this asymptotic may not be enough for the obtaining good enhancement factor.

\section{Discussion and Conclusion}

Low error on the validation data set indicates that the developed approach to the architecture of the XC functional interpolates well the existing XC functionals and has the high generalizing ability. Furthermore, in the framework of the proposed architecture, one can train NN-E and NN-V separately, making it flexible to use. The basic strategy to create working functional in the framework of the proposed architecture would be initially to train the NN-V part on a specific type of existing XC functional such as LDA, GGA, or meta-GGA, using various types of input features. Then the NN-E part of the neural network is trained on the data obtained by accurate post-Hartree-Fock or quantum Monte Carlo methods.


The main advantage of the NN approach in comparison with other interpolation techniques for XC functionals is its flexibility to incorporate exchange-correlation data from different sources, such as post-Hartree-Fock and quantum Monte Carlo. It is possible that application of the NN to interpolate high-level XC quantum data could eliminate many heuristics used in the traditional construction of XC functionals.



\bibliography{apssamp}

\providecommand{\noopsort}[1]{}\providecommand{\singleletter}[1]{#1}%
\begin{thebibliography}{33}%
\makeatletter
\providecommand \@ifxundefined [1]{%
 \@ifx{#1\undefined}
}%
\providecommand \@ifnum [1]{%
 \ifnum #1\expandafter \@firstoftwo
 \else \expandafter \@secondoftwo
 \fi
}%
\providecommand \@ifx [1]{%
 \ifx #1\expandafter \@firstoftwo
 \else \expandafter \@secondoftwo
 \fi
}%
\providecommand \natexlab [1]{#1}%
\providecommand \enquote  [1]{``#1''}%
\providecommand \bibnamefont  [1]{#1}%
\providecommand \bibfnamefont [1]{#1}%
\providecommand \citenamefont [1]{#1}%
\providecommand \href@noop [0]{\@secondoftwo}%
\providecommand \href [0]{\begingroup \@sanitize@url \@href}%
\providecommand \@href[1]{\@@startlink{#1}\@@href}%
\providecommand \@@href[1]{\endgroup#1\@@endlink}%
\providecommand \@sanitize@url [0]{\catcode `\\12\catcode `\$12\catcode
  `\&12\catcode `\#12\catcode `\^12\catcode `\_12\catcode `\%12\relax}%
\providecommand \@@startlink[1]{}%
\providecommand \@@endlink[0]{}%
\providecommand \url  [0]{\begingroup\@sanitize@url \@url }%
\providecommand \@url [1]{\endgroup\@href {#1}{\urlprefix }}%
\providecommand \urlprefix  [0]{URL }%
\providecommand \Eprint [0]{\href }%
\providecommand \doibase [0]{https://doi.org/}%
\providecommand \selectlanguage [0]{\@gobble}%
\providecommand \bibinfo  [0]{\@secondoftwo}%
\providecommand \bibfield  [0]{\@secondoftwo}%
\providecommand \translation [1]{[#1]}%
\providecommand \BibitemOpen [0]{}%
\providecommand \bibitemStop [0]{}%
\providecommand \bibitemNoStop [0]{.\EOS\space}%
\providecommand \EOS [0]{\spacefactor3000\relax}%
\providecommand \BibitemShut  [1]{\csname bibitem#1\endcsname}%
\let\auto@bib@innerbib\@empty
\bibitem [{\citenamefont {Hohenberg}\ and\ \citenamefont
  {Kohn}(1964)}]{hohenberg1964inhomogeneous}%
  \BibitemOpen
  \bibfield  {author} {\bibinfo {author} {\bibfnamefont {P.}~\bibnamefont
  {Hohenberg}}\ and\ \bibinfo {author} {\bibfnamefont {W.}~\bibnamefont
  {Kohn}},\ }\bibfield  {title} {\bibinfo {title} {Inhomogeneous electron
  gas},\ }\href@noop {} {\bibfield  {journal} {\bibinfo  {journal} {Physical
  review}\ }\textbf {\bibinfo {volume} {136}},\ \bibinfo {pages} {B864}
  (\bibinfo {year} {1964})}\BibitemShut {NoStop}%
\bibitem [{\citenamefont {Kohn}\ and\ \citenamefont
  {Sham}(1965)}]{kohn1965self}%
  \BibitemOpen
  \bibfield  {author} {\bibinfo {author} {\bibfnamefont {W.}~\bibnamefont
  {Kohn}}\ and\ \bibinfo {author} {\bibfnamefont {L.~J.}\ \bibnamefont
  {Sham}},\ }\bibfield  {title} {\bibinfo {title} {Self-consistent equations
  including exchange and correlation effects},\ }\href@noop {} {\bibfield
  {journal} {\bibinfo  {journal} {Physical review}\ }\textbf {\bibinfo {volume}
  {140}},\ \bibinfo {pages} {A1133} (\bibinfo {year} {1965})}\BibitemShut
  {NoStop}%
\bibitem [{\citenamefont {Ceperley}\ and\ \citenamefont
  {Alder}(1980)}]{ceperley1980ground}%
  \BibitemOpen
  \bibfield  {author} {\bibinfo {author} {\bibfnamefont {D.~M.}\ \bibnamefont
  {Ceperley}}\ and\ \bibinfo {author} {\bibfnamefont {B.}~\bibnamefont
  {Alder}},\ }\bibfield  {title} {\bibinfo {title} {Ground state of the
  electron gas by a stochastic method},\ }\href@noop {} {\bibfield  {journal}
  {\bibinfo  {journal} {Physical Review Letters}\ }\textbf {\bibinfo {volume}
  {45}},\ \bibinfo {pages} {566} (\bibinfo {year} {1980})}\BibitemShut
  {NoStop}%
\bibitem [{\citenamefont {Zhao}\ \emph {et~al.}(2005)\citenamefont {Zhao},
  \citenamefont {Schultz},\ and\ \citenamefont {Truhlar}}]{zhao2005exchange}%
  \BibitemOpen
  \bibfield  {author} {\bibinfo {author} {\bibfnamefont {Y.}~\bibnamefont
  {Zhao}}, \bibinfo {author} {\bibfnamefont {N.~E.}\ \bibnamefont {Schultz}},\
  and\ \bibinfo {author} {\bibfnamefont {D.~G.}\ \bibnamefont {Truhlar}},\
  }\href@noop {} {\bibinfo {title} {Exchange-correlation functional with broad
  accuracy for metallic and nonmetallic compounds, kinetics, and noncovalent
  interactions}} (\bibinfo {year} {2005})\BibitemShut {NoStop}%
\bibitem [{\citenamefont {Vosko}\ \emph {et~al.}(1980)\citenamefont {Vosko},
  \citenamefont {Wilk},\ and\ \citenamefont {Nusair}}]{vosko1980accurate}%
  \BibitemOpen
  \bibfield  {author} {\bibinfo {author} {\bibfnamefont {S.~H.}\ \bibnamefont
  {Vosko}}, \bibinfo {author} {\bibfnamefont {L.}~\bibnamefont {Wilk}},\ and\
  \bibinfo {author} {\bibfnamefont {M.}~\bibnamefont {Nusair}},\ }\bibfield
  {title} {\bibinfo {title} {Accurate spin-dependent electron liquid
  correlation energies for local spin density calculations: a critical
  analysis},\ }\href@noop {} {\bibfield  {journal} {\bibinfo  {journal}
  {Canadian Journal of physics}\ }\textbf {\bibinfo {volume} {58}},\ \bibinfo
  {pages} {1200} (\bibinfo {year} {1980})}\BibitemShut {NoStop}%
\bibitem [{\citenamefont {Perdew}\ and\ \citenamefont
  {Zunger}(1981)}]{perdew1981self}%
  \BibitemOpen
  \bibfield  {author} {\bibinfo {author} {\bibfnamefont {J.~P.}\ \bibnamefont
  {Perdew}}\ and\ \bibinfo {author} {\bibfnamefont {A.}~\bibnamefont
  {Zunger}},\ }\bibfield  {title} {\bibinfo {title} {Self-interaction
  correction to density-functional approximations for many-electron systems},\
  }\href@noop {} {\bibfield  {journal} {\bibinfo  {journal} {Physical Review
  B}\ }\textbf {\bibinfo {volume} {23}},\ \bibinfo {pages} {5048} (\bibinfo
  {year} {1981})}\BibitemShut {NoStop}%
\bibitem [{\citenamefont {Perdew}\ and\ \citenamefont
  {Wang}(1992)}]{perdew1992accurate}%
  \BibitemOpen
  \bibfield  {author} {\bibinfo {author} {\bibfnamefont {J.~P.}\ \bibnamefont
  {Perdew}}\ and\ \bibinfo {author} {\bibfnamefont {Y.}~\bibnamefont {Wang}},\
  }\bibfield  {title} {\bibinfo {title} {Accurate and simple analytic
  representation of the electron-gas correlation energy},\ }\href@noop {}
  {\bibfield  {journal} {\bibinfo  {journal} {Physical Review B}\ }\textbf
  {\bibinfo {volume} {45}},\ \bibinfo {pages} {13244} (\bibinfo {year}
  {1992})}\BibitemShut {NoStop}%
\bibitem [{\citenamefont {Wang}\ and\ \citenamefont
  {Perdew}(1991)}]{wang1991spin}%
  \BibitemOpen
  \bibfield  {author} {\bibinfo {author} {\bibfnamefont {Y.}~\bibnamefont
  {Wang}}\ and\ \bibinfo {author} {\bibfnamefont {J.~P.}\ \bibnamefont
  {Perdew}},\ }\bibfield  {title} {\bibinfo {title} {Spin scaling of the
  electron-gas correlation energy in the high-density limit},\ }\href@noop {}
  {\bibfield  {journal} {\bibinfo  {journal} {Physical Review B}\ }\textbf
  {\bibinfo {volume} {43}},\ \bibinfo {pages} {8911} (\bibinfo {year}
  {1991})}\BibitemShut {NoStop}%
\bibitem [{\citenamefont {Perdew}\ \emph {et~al.}(1992)\citenamefont {Perdew},
  \citenamefont {Chevary}, \citenamefont {Vosko}, \citenamefont {Jackson},
  \citenamefont {Pederson}, \citenamefont {Singh},\ and\ \citenamefont
  {Fiolhais}}]{perdew1992atoms}%
  \BibitemOpen
  \bibfield  {author} {\bibinfo {author} {\bibfnamefont {J.~P.}\ \bibnamefont
  {Perdew}}, \bibinfo {author} {\bibfnamefont {J.~A.}\ \bibnamefont {Chevary}},
  \bibinfo {author} {\bibfnamefont {S.~H.}\ \bibnamefont {Vosko}}, \bibinfo
  {author} {\bibfnamefont {K.~A.}\ \bibnamefont {Jackson}}, \bibinfo {author}
  {\bibfnamefont {M.~R.}\ \bibnamefont {Pederson}}, \bibinfo {author}
  {\bibfnamefont {D.~J.}\ \bibnamefont {Singh}},\ and\ \bibinfo {author}
  {\bibfnamefont {C.}~\bibnamefont {Fiolhais}},\ }\bibfield  {title} {\bibinfo
  {title} {Atoms, molecules, solids, and surfaces: Applications of the
  generalized gradient approximation for exchange and correlation},\
  }\href@noop {} {\bibfield  {journal} {\bibinfo  {journal} {Physical review
  B}\ }\textbf {\bibinfo {volume} {46}},\ \bibinfo {pages} {6671} (\bibinfo
  {year} {1992})}\BibitemShut {NoStop}%
\bibitem [{\citenamefont {Perdew}\ \emph {et~al.}(1996)\citenamefont {Perdew},
  \citenamefont {Burke},\ and\ \citenamefont
  {Ernzerhof}}]{perdew1996generalized}%
  \BibitemOpen
  \bibfield  {author} {\bibinfo {author} {\bibfnamefont {J.~P.}\ \bibnamefont
  {Perdew}}, \bibinfo {author} {\bibfnamefont {K.}~\bibnamefont {Burke}},\ and\
  \bibinfo {author} {\bibfnamefont {M.}~\bibnamefont {Ernzerhof}},\ }\bibfield
  {title} {\bibinfo {title} {Generalized gradient approximation made simple},\
  }\href@noop {} {\bibfield  {journal} {\bibinfo  {journal} {Physical review
  letters}\ }\textbf {\bibinfo {volume} {77}},\ \bibinfo {pages} {3865}
  (\bibinfo {year} {1996})}\BibitemShut {NoStop}%
\bibitem [{\citenamefont {Mardirossian}\ and\ \citenamefont
  {Head-Gordon}(2017)}]{mardirossian2017thirty}%
  \BibitemOpen
  \bibfield  {author} {\bibinfo {author} {\bibfnamefont {N.}~\bibnamefont
  {Mardirossian}}\ and\ \bibinfo {author} {\bibfnamefont {M.}~\bibnamefont
  {Head-Gordon}},\ }\bibfield  {title} {\bibinfo {title} {Thirty years of
  density functional theory in computational chemistry: an overview and
  extensive assessment of 200 density functionals},\ }\href@noop {} {\bibfield
  {journal} {\bibinfo  {journal} {Molecular Physics}\ }\textbf {\bibinfo
  {volume} {115}},\ \bibinfo {pages} {2315} (\bibinfo {year}
  {2017})}\BibitemShut {NoStop}%
\bibitem [{\citenamefont {Lani}\ \emph {et~al.}(2016)\citenamefont {Lani},
  \citenamefont {Di~Marino}, \citenamefont {Gerolin}, \citenamefont {van
  Leeuwen},\ and\ \citenamefont {Gori-Giorgi}}]{lani2016adiabatic}%
  \BibitemOpen
  \bibfield  {author} {\bibinfo {author} {\bibfnamefont {G.}~\bibnamefont
  {Lani}}, \bibinfo {author} {\bibfnamefont {S.}~\bibnamefont {Di~Marino}},
  \bibinfo {author} {\bibfnamefont {A.}~\bibnamefont {Gerolin}}, \bibinfo
  {author} {\bibfnamefont {R.}~\bibnamefont {van Leeuwen}},\ and\ \bibinfo
  {author} {\bibfnamefont {P.}~\bibnamefont {Gori-Giorgi}},\ }\bibfield
  {title} {\bibinfo {title} {The adiabatic strictly-correlated-electrons
  functional: kernel and exact properties},\ }\href@noop {} {\bibfield
  {journal} {\bibinfo  {journal} {Physical Chemistry Chemical Physics}\
  }\textbf {\bibinfo {volume} {18}},\ \bibinfo {pages} {21092} (\bibinfo {year}
  {2016})}\BibitemShut {NoStop}%
\bibitem [{\citenamefont {Maier}\ \emph {et~al.}(2016)\citenamefont {Maier},
  \citenamefont {Haasler}, \citenamefont {Arbuznikov},\ and\ \citenamefont
  {Kaupp}}]{maier2016new}%
  \BibitemOpen
  \bibfield  {author} {\bibinfo {author} {\bibfnamefont {T.~M.}\ \bibnamefont
  {Maier}}, \bibinfo {author} {\bibfnamefont {M.}~\bibnamefont {Haasler}},
  \bibinfo {author} {\bibfnamefont {A.~V.}\ \bibnamefont {Arbuznikov}},\ and\
  \bibinfo {author} {\bibfnamefont {M.}~\bibnamefont {Kaupp}},\ }\bibfield
  {title} {\bibinfo {title} {New approaches for the calibration of
  exchange-energy densities in local hybrid functionals},\ }\href@noop {}
  {\bibfield  {journal} {\bibinfo  {journal} {Physical Chemistry Chemical
  Physics}\ }\textbf {\bibinfo {volume} {18}},\ \bibinfo {pages} {21133}
  (\bibinfo {year} {2016})}\BibitemShut {NoStop}%
\bibitem [{\citenamefont {Mori-S{\'a}nchez}\ and\ \citenamefont
  {Cohen}(2014)}]{mori2014derivative}%
  \BibitemOpen
  \bibfield  {author} {\bibinfo {author} {\bibfnamefont {P.}~\bibnamefont
  {Mori-S{\'a}nchez}}\ and\ \bibinfo {author} {\bibfnamefont {A.~J.}\
  \bibnamefont {Cohen}},\ }\bibfield  {title} {\bibinfo {title} {The derivative
  discontinuity of the exchange--correlation functional},\ }\href@noop {}
  {\bibfield  {journal} {\bibinfo  {journal} {Physical Chemistry Chemical
  Physics}\ }\textbf {\bibinfo {volume} {16}},\ \bibinfo {pages} {14378}
  (\bibinfo {year} {2014})}\BibitemShut {NoStop}%
\bibitem [{\citenamefont {Mori-S{\'a}nchez}\ and\ \citenamefont
  {Cohen}(2018)}]{mori2018exact}%
  \BibitemOpen
  \bibfield  {author} {\bibinfo {author} {\bibfnamefont {P.}~\bibnamefont
  {Mori-S{\'a}nchez}}\ and\ \bibinfo {author} {\bibfnamefont {A.~J.}\
  \bibnamefont {Cohen}},\ }\bibfield  {title} {\bibinfo {title} {Exact density
  functional obtained via the levy constrained search},\ }\href@noop {}
  {\bibfield  {journal} {\bibinfo  {journal} {The journal of physical chemistry
  letters}\ }\textbf {\bibinfo {volume} {9}},\ \bibinfo {pages} {4910}
  (\bibinfo {year} {2018})}\BibitemShut {NoStop}%
\bibitem [{\citenamefont {Needs}\ \emph {et~al.}(2009)\citenamefont {Needs},
  \citenamefont {Towler}, \citenamefont {Drummond},\ and\ \citenamefont
  {R{\'\i}os}}]{needs2009continuum}%
  \BibitemOpen
  \bibfield  {author} {\bibinfo {author} {\bibfnamefont {R.}~\bibnamefont
  {Needs}}, \bibinfo {author} {\bibfnamefont {M.}~\bibnamefont {Towler}},
  \bibinfo {author} {\bibfnamefont {N.}~\bibnamefont {Drummond}},\ and\
  \bibinfo {author} {\bibfnamefont {P.~L.}\ \bibnamefont {R{\'\i}os}},\
  }\bibfield  {title} {\bibinfo {title} {Continuum variational and diffusion
  quantum monte carlo calculations},\ }\href@noop {} {\bibfield  {journal}
  {\bibinfo  {journal} {Journal of Physics: Condensed Matter}\ }\textbf
  {\bibinfo {volume} {22}},\ \bibinfo {pages} {023201} (\bibinfo {year}
  {2009})}\BibitemShut {NoStop}%
\bibitem [{\citenamefont {Koloren{\v{c}}}\ and\ \citenamefont
  {Mitas}(2011)}]{kolorenvc2011applications}%
  \BibitemOpen
  \bibfield  {author} {\bibinfo {author} {\bibfnamefont {J.}~\bibnamefont
  {Koloren{\v{c}}}}\ and\ \bibinfo {author} {\bibfnamefont {L.}~\bibnamefont
  {Mitas}},\ }\bibfield  {title} {\bibinfo {title} {Applications of quantum
  monte carlo methods in condensed systems},\ }\href@noop {} {\bibfield
  {journal} {\bibinfo  {journal} {Reports on Progress in Physics}\ }\textbf
  {\bibinfo {volume} {74}},\ \bibinfo {pages} {026502} (\bibinfo {year}
  {2011})}\BibitemShut {NoStop}%
\bibitem [{\citenamefont {Cremer}(2011)}]{cremer2011moller}%
  \BibitemOpen
  \bibfield  {author} {\bibinfo {author} {\bibfnamefont {D.}~\bibnamefont
  {Cremer}},\ }\bibfield  {title} {\bibinfo {title} {M{\o}ller--plesset
  perturbation theory: from small molecule methods to methods for thousands of
  atoms},\ }\href@noop {} {\bibfield  {journal} {\bibinfo  {journal} {Wiley
  Interdisciplinary Reviews: Computational Molecular Science}\ }\textbf
  {\bibinfo {volume} {1}},\ \bibinfo {pages} {509} (\bibinfo {year}
  {2011})}\BibitemShut {NoStop}%
\bibitem [{\citenamefont {Cybenko}(1989)}]{cybenko1989approximation}%
  \BibitemOpen
  \bibfield  {author} {\bibinfo {author} {\bibfnamefont {G.}~\bibnamefont
  {Cybenko}},\ }\bibfield  {title} {\bibinfo {title} {Approximation by
  superpositions of a sigmoidal function},\ }\href@noop {} {\bibfield
  {journal} {\bibinfo  {journal} {Mathematics of control, signals and systems}\
  }\textbf {\bibinfo {volume} {2}},\ \bibinfo {pages} {303} (\bibinfo {year}
  {1989})}\BibitemShut {NoStop}%
\bibitem [{\citenamefont {Tozer}\ \emph {et~al.}(1996)\citenamefont {Tozer},
  \citenamefont {Ingamells},\ and\ \citenamefont {Handy}}]{tozer1996exchange}%
  \BibitemOpen
  \bibfield  {author} {\bibinfo {author} {\bibfnamefont {D.~J.}\ \bibnamefont
  {Tozer}}, \bibinfo {author} {\bibfnamefont {V.~E.}\ \bibnamefont
  {Ingamells}},\ and\ \bibinfo {author} {\bibfnamefont {N.~C.}\ \bibnamefont
  {Handy}},\ }\bibfield  {title} {\bibinfo {title} {Exchange-correlation
  potentials},\ }\href@noop {} {\bibfield  {journal} {\bibinfo  {journal} {The
  Journal of chemical physics}\ }\textbf {\bibinfo {volume} {105}},\ \bibinfo
  {pages} {9200} (\bibinfo {year} {1996})}\BibitemShut {NoStop}%
\bibitem [{\citenamefont {Nagai}\ \emph {et~al.}(2018)\citenamefont {Nagai},
  \citenamefont {Akashi}, \citenamefont {Sasaki},\ and\ \citenamefont
  {Tsuneyuki}}]{nagai2018neural}%
  \BibitemOpen
  \bibfield  {author} {\bibinfo {author} {\bibfnamefont {R.}~\bibnamefont
  {Nagai}}, \bibinfo {author} {\bibfnamefont {R.}~\bibnamefont {Akashi}},
  \bibinfo {author} {\bibfnamefont {S.}~\bibnamefont {Sasaki}},\ and\ \bibinfo
  {author} {\bibfnamefont {S.}~\bibnamefont {Tsuneyuki}},\ }\bibfield  {title}
  {\bibinfo {title} {Neural-network kohn-sham exchange-correlation potential
  and its out-of-training transferability},\ }\href@noop {} {\bibfield
  {journal} {\bibinfo  {journal} {The Journal of chemical physics}\ }\textbf
  {\bibinfo {volume} {148}},\ \bibinfo {pages} {241737} (\bibinfo {year}
  {2018})}\BibitemShut {NoStop}%
\bibitem [{\citenamefont {Nagai}\ \emph {et~al.}(2020)\citenamefont {Nagai},
  \citenamefont {Akashi},\ and\ \citenamefont {Sugino}}]{nagai2020completing}%
  \BibitemOpen
  \bibfield  {author} {\bibinfo {author} {\bibfnamefont {R.}~\bibnamefont
  {Nagai}}, \bibinfo {author} {\bibfnamefont {R.}~\bibnamefont {Akashi}},\ and\
  \bibinfo {author} {\bibfnamefont {O.}~\bibnamefont {Sugino}},\ }\bibfield
  {title} {\bibinfo {title} {Completing density functional theory by machine
  learning hidden messages from molecules},\ }\href@noop {} {\bibfield
  {journal} {\bibinfo  {journal} {npj Computational Materials}\ }\textbf
  {\bibinfo {volume} {6}},\ \bibinfo {pages} {1} (\bibinfo {year}
  {2020})}\BibitemShut {NoStop}%
\bibitem [{\citenamefont {Lei}\ and\ \citenamefont
  {Medford}(2019)}]{lei2019design}%
  \BibitemOpen
  \bibfield  {author} {\bibinfo {author} {\bibfnamefont {X.}~\bibnamefont
  {Lei}}\ and\ \bibinfo {author} {\bibfnamefont {A.~J.}\ \bibnamefont
  {Medford}},\ }\bibfield  {title} {\bibinfo {title} {Design and analysis of
  machine learning exchange-correlation functionals via rotationally invariant
  convolutional descriptors},\ }\href@noop {} {\bibfield  {journal} {\bibinfo
  {journal} {Physical Review Materials}\ }\textbf {\bibinfo {volume} {3}},\
  \bibinfo {pages} {063801} (\bibinfo {year} {2019})}\BibitemShut {NoStop}%
\bibitem [{\citenamefont {Ramos}\ and\ \citenamefont
  {Pavanello}(2019)}]{ramos2019static}%
  \BibitemOpen
  \bibfield  {author} {\bibinfo {author} {\bibfnamefont {P.}~\bibnamefont
  {Ramos}}\ and\ \bibinfo {author} {\bibfnamefont {M.}~\bibnamefont
  {Pavanello}},\ }\bibfield  {title} {\bibinfo {title} {Static correlation
  density functional theory},\ }\href@noop {} {\bibfield  {journal} {\bibinfo
  {journal} {arXiv preprint arXiv:1906.06661}\ } (\bibinfo {year}
  {2019})}\BibitemShut {NoStop}%
\bibitem [{\citenamefont {Ryabov}\ \emph {et~al.}(2020)\citenamefont {Ryabov},
  \citenamefont {Akhatov},\ and\ \citenamefont {Zhilyaev}}]{ryabov2020neural}%
  \BibitemOpen
  \bibfield  {author} {\bibinfo {author} {\bibfnamefont {A.}~\bibnamefont
  {Ryabov}}, \bibinfo {author} {\bibfnamefont {I.}~\bibnamefont {Akhatov}},\
  and\ \bibinfo {author} {\bibfnamefont {P.}~\bibnamefont {Zhilyaev}},\
  }\bibfield  {title} {\bibinfo {title} {Neural network interpolation of
  exchange-correlation functional},\ }\href@noop {} {\bibfield  {journal}
  {\bibinfo  {journal} {Scientific reports}\ }\textbf {\bibinfo {volume}
  {10}},\ \bibinfo {pages} {1} (\bibinfo {year} {2020})}\BibitemShut {NoStop}%
\bibitem [{\citenamefont {Li}\ \emph {et~al.}(2021)\citenamefont {Li},
  \citenamefont {Hoyer}, \citenamefont {Pederson}, \citenamefont {Sun},
  \citenamefont {Cubuk}, \citenamefont {Riley}, \citenamefont {Burke} \emph
  {et~al.}}]{li2021kohn}%
  \BibitemOpen
  \bibfield  {author} {\bibinfo {author} {\bibfnamefont {L.}~\bibnamefont
  {Li}}, \bibinfo {author} {\bibfnamefont {S.}~\bibnamefont {Hoyer}}, \bibinfo
  {author} {\bibfnamefont {R.}~\bibnamefont {Pederson}}, \bibinfo {author}
  {\bibfnamefont {R.}~\bibnamefont {Sun}}, \bibinfo {author} {\bibfnamefont
  {E.~D.}\ \bibnamefont {Cubuk}}, \bibinfo {author} {\bibfnamefont
  {P.}~\bibnamefont {Riley}}, \bibinfo {author} {\bibfnamefont
  {K.}~\bibnamefont {Burke}}, \emph {et~al.},\ }\bibfield  {title} {\bibinfo
  {title} {Kohn-sham equations as regularizer: Building prior knowledge into
  machine-learned physics},\ }\href@noop {} {\bibfield  {journal} {\bibinfo
  {journal} {Physical Review Letters}\ }\textbf {\bibinfo {volume} {126}},\
  \bibinfo {pages} {036401} (\bibinfo {year} {2021})}\BibitemShut {NoStop}%
\bibitem [{\citenamefont {Perdew}\ \emph {et~al.}(1997)\citenamefont {Perdew},
  \citenamefont {Burke},\ and\ \citenamefont
  {Ernzerhof}}]{PhysRevLett.78.1396}%
  \BibitemOpen
  \bibfield  {author} {\bibinfo {author} {\bibfnamefont {J.~P.}\ \bibnamefont
  {Perdew}}, \bibinfo {author} {\bibfnamefont {K.}~\bibnamefont {Burke}},\ and\
  \bibinfo {author} {\bibfnamefont {M.}~\bibnamefont {Ernzerhof}},\ }\bibfield
  {title} {\bibinfo {title} {Generalized gradient approximation made simple
  [phys. rev. lett. 77, 3865 (1996)]},\ }\href@noop {} {\bibfield  {journal}
  {\bibinfo  {journal} {Phys. Rev. Lett.}\ }\textbf {\bibinfo {volume} {78}},\
  \bibinfo {pages} {1396} (\bibinfo {year} {1997})}\BibitemShut {NoStop}%
\bibitem [{\citenamefont {Andrade}\ \emph {et~al.}(2015)\citenamefont
  {Andrade}, \citenamefont {Strubbe}, \citenamefont {De~Giovannini},
  \citenamefont {Larsen}, \citenamefont {Oliveira}, \citenamefont
  {Alberdi-Rodriguez}, \citenamefont {Varas}, \citenamefont {Theophilou},
  \citenamefont {Helbig}, \citenamefont {Verstraete} \emph
  {et~al.}}]{andrade2015real}%
  \BibitemOpen
  \bibfield  {author} {\bibinfo {author} {\bibfnamefont {X.}~\bibnamefont
  {Andrade}}, \bibinfo {author} {\bibfnamefont {D.}~\bibnamefont {Strubbe}},
  \bibinfo {author} {\bibfnamefont {U.}~\bibnamefont {De~Giovannini}}, \bibinfo
  {author} {\bibfnamefont {A.~H.}\ \bibnamefont {Larsen}}, \bibinfo {author}
  {\bibfnamefont {M.~J.}\ \bibnamefont {Oliveira}}, \bibinfo {author}
  {\bibfnamefont {J.}~\bibnamefont {Alberdi-Rodriguez}}, \bibinfo {author}
  {\bibfnamefont {A.}~\bibnamefont {Varas}}, \bibinfo {author} {\bibfnamefont
  {I.}~\bibnamefont {Theophilou}}, \bibinfo {author} {\bibfnamefont
  {N.}~\bibnamefont {Helbig}}, \bibinfo {author} {\bibfnamefont {M.~J.}\
  \bibnamefont {Verstraete}}, \emph {et~al.},\ }\bibfield  {title} {\bibinfo
  {title} {Real-space grids and the octopus code as tools for the development
  of new simulation approaches for electronic systems},\ }\href@noop {}
  {\bibfield  {journal} {\bibinfo  {journal} {Physical Chemistry Chemical
  Physics}\ }\textbf {\bibinfo {volume} {17}},\ \bibinfo {pages} {31371}
  (\bibinfo {year} {2015})}\BibitemShut {NoStop}%
\bibitem [{\citenamefont {Andrade}\ \emph {et~al.}(2012)\citenamefont
  {Andrade}, \citenamefont {Alberdi-Rodriguez}, \citenamefont {Strubbe},
  \citenamefont {Oliveira}, \citenamefont {Nogueira}, \citenamefont {Castro},
  \citenamefont {Muguerza}, \citenamefont {Arruabarrena}, \citenamefont
  {Louie}, \citenamefont {Aspuru-Guzik} \emph {et~al.}}]{andrade2012time}%
  \BibitemOpen
  \bibfield  {author} {\bibinfo {author} {\bibfnamefont {X.}~\bibnamefont
  {Andrade}}, \bibinfo {author} {\bibfnamefont {J.}~\bibnamefont
  {Alberdi-Rodriguez}}, \bibinfo {author} {\bibfnamefont {D.~A.}\ \bibnamefont
  {Strubbe}}, \bibinfo {author} {\bibfnamefont {M.~J.}\ \bibnamefont
  {Oliveira}}, \bibinfo {author} {\bibfnamefont {F.}~\bibnamefont {Nogueira}},
  \bibinfo {author} {\bibfnamefont {A.}~\bibnamefont {Castro}}, \bibinfo
  {author} {\bibfnamefont {J.}~\bibnamefont {Muguerza}}, \bibinfo {author}
  {\bibfnamefont {A.}~\bibnamefont {Arruabarrena}}, \bibinfo {author}
  {\bibfnamefont {S.~G.}\ \bibnamefont {Louie}}, \bibinfo {author}
  {\bibfnamefont {A.}~\bibnamefont {Aspuru-Guzik}}, \emph {et~al.},\ }\bibfield
   {title} {\bibinfo {title} {Time-dependent density-functional theory in
  massively parallel computer architectures: the octopus project},\ }\href@noop
  {} {\bibfield  {journal} {\bibinfo  {journal} {Journal of Physics: Condensed
  Matter}\ }\textbf {\bibinfo {volume} {24}},\ \bibinfo {pages} {233202}
  (\bibinfo {year} {2012})}\BibitemShut {NoStop}%
\bibitem [{\citenamefont {Andrade}\ and\ \citenamefont
  {Aspuru-Guzik}(2013)}]{andrade2013real}%
  \BibitemOpen
  \bibfield  {author} {\bibinfo {author} {\bibfnamefont {X.}~\bibnamefont
  {Andrade}}\ and\ \bibinfo {author} {\bibfnamefont {A.}~\bibnamefont
  {Aspuru-Guzik}},\ }\bibfield  {title} {\bibinfo {title} {Real-space density
  functional theory on graphical processing units: computational approach and
  comparison to gaussian basis set methods},\ }\href@noop {} {\bibfield
  {journal} {\bibinfo  {journal} {Journal of chemical theory and computation}\
  }\textbf {\bibinfo {volume} {9}},\ \bibinfo {pages} {4360} (\bibinfo {year}
  {2013})}\BibitemShut {NoStop}%
\bibitem [{\citenamefont {Lehtola}\ \emph {et~al.}(2018)\citenamefont
  {Lehtola}, \citenamefont {Steigemann}, \citenamefont {Oliveira},\ and\
  \citenamefont {Marques}}]{lehtola2018recent}%
  \BibitemOpen
  \bibfield  {author} {\bibinfo {author} {\bibfnamefont {S.}~\bibnamefont
  {Lehtola}}, \bibinfo {author} {\bibfnamefont {C.}~\bibnamefont {Steigemann}},
  \bibinfo {author} {\bibfnamefont {M.~J.}\ \bibnamefont {Oliveira}},\ and\
  \bibinfo {author} {\bibfnamefont {M.~A.}\ \bibnamefont {Marques}},\
  }\bibfield  {title} {\bibinfo {title} {Recent developments in libxc—a
  comprehensive library of functionals for density functional theory},\
  }\href@noop {} {\bibfield  {journal} {\bibinfo  {journal} {SoftwareX}\
  }\textbf {\bibinfo {volume} {7}},\ \bibinfo {pages} {1} (\bibinfo {year}
  {2018})}\BibitemShut {NoStop}%
\bibitem [{\citenamefont {Paszke}\ \emph {et~al.}(2019)\citenamefont {Paszke},
  \citenamefont {Gross}, \citenamefont {Massa}, \citenamefont {Lerer},
  \citenamefont {Bradbury}, \citenamefont {Chanan}, \citenamefont {Killeen},
  \citenamefont {Lin}, \citenamefont {Gimelshein}, \citenamefont {Antiga},
  \citenamefont {Desmaison}, \citenamefont {Kopf}, \citenamefont {Yang},
  \citenamefont {DeVito}, \citenamefont {Raison}, \citenamefont {Tejani},
  \citenamefont {Chilamkurthy}, \citenamefont {Steiner}, \citenamefont {Fang},
  \citenamefont {Bai},\ and\ \citenamefont {Chintala}}]{NEURIPS2019_9015}%
  \BibitemOpen
  \bibfield  {author} {\bibinfo {author} {\bibfnamefont {A.}~\bibnamefont
  {Paszke}}, \bibinfo {author} {\bibfnamefont {S.}~\bibnamefont {Gross}},
  \bibinfo {author} {\bibfnamefont {F.}~\bibnamefont {Massa}}, \bibinfo
  {author} {\bibfnamefont {A.}~\bibnamefont {Lerer}}, \bibinfo {author}
  {\bibfnamefont {J.}~\bibnamefont {Bradbury}}, \bibinfo {author}
  {\bibfnamefont {G.}~\bibnamefont {Chanan}}, \bibinfo {author} {\bibfnamefont
  {T.}~\bibnamefont {Killeen}}, \bibinfo {author} {\bibfnamefont
  {Z.}~\bibnamefont {Lin}}, \bibinfo {author} {\bibfnamefont {N.}~\bibnamefont
  {Gimelshein}}, \bibinfo {author} {\bibfnamefont {L.}~\bibnamefont {Antiga}},
  \bibinfo {author} {\bibfnamefont {A.}~\bibnamefont {Desmaison}}, \bibinfo
  {author} {\bibfnamefont {A.}~\bibnamefont {Kopf}}, \bibinfo {author}
  {\bibfnamefont {E.}~\bibnamefont {Yang}}, \bibinfo {author} {\bibfnamefont
  {Z.}~\bibnamefont {DeVito}}, \bibinfo {author} {\bibfnamefont
  {M.}~\bibnamefont {Raison}}, \bibinfo {author} {\bibfnamefont
  {A.}~\bibnamefont {Tejani}}, \bibinfo {author} {\bibfnamefont
  {S.}~\bibnamefont {Chilamkurthy}}, \bibinfo {author} {\bibfnamefont
  {B.}~\bibnamefont {Steiner}}, \bibinfo {author} {\bibfnamefont
  {L.}~\bibnamefont {Fang}}, \bibinfo {author} {\bibfnamefont {J.}~\bibnamefont
  {Bai}},\ and\ \bibinfo {author} {\bibfnamefont {S.}~\bibnamefont
  {Chintala}},\ }\bibfield  {title} {\bibinfo {title} {Pytorch: An imperative
  style, high-performance deep learning library},\ }in\ \href
  {http://papers.neurips.cc/paper/9015-pytorch-an-imperative-style-high-performance-deep-learning-library.pdf}
  {\emph {\bibinfo {booktitle} {Advances in Neural Information Processing
  Systems 32}}},\ \bibinfo {editor} {edited by\ \bibinfo {editor}
  {\bibfnamefont {H.}~\bibnamefont {Wallach}}, \bibinfo {editor} {\bibfnamefont
  {H.}~\bibnamefont {Larochelle}}, \bibinfo {editor} {\bibfnamefont
  {A.}~\bibnamefont {Beygelzimer}}, \bibinfo {editor} {\bibfnamefont
  {F.}~\bibnamefont {d\textquotesingle Alch\'{e}-Buc}}, \bibinfo {editor}
  {\bibfnamefont {E.}~\bibnamefont {Fox}},\ and\ \bibinfo {editor}
  {\bibfnamefont {R.}~\bibnamefont {Garnett}}}\ (\bibinfo  {publisher} {Curran
  Associates, Inc.},\ \bibinfo {year} {2019})\ pp.\ \bibinfo {pages}
  {8024--8035}\BibitemShut {NoStop}%
\bibitem [{\citenamefont {Clevert}\ \emph {et~al.}(2016)\citenamefont
  {Clevert}, \citenamefont {Unterthiner},\ and\ \citenamefont
  {Hochreiter}}]{clevert2016fast}%
  \BibitemOpen
  \bibfield  {author} {\bibinfo {author} {\bibfnamefont {D.-A.}\ \bibnamefont
  {Clevert}}, \bibinfo {author} {\bibfnamefont {T.}~\bibnamefont
  {Unterthiner}},\ and\ \bibinfo {author} {\bibfnamefont {S.}~\bibnamefont
  {Hochreiter}},\ }\href@noop {} {\bibinfo {title} {Fast and accurate deep
  network learning by exponential linear units (elus)}} (\bibinfo {year}
  {2016}),\ \Eprint {https://arxiv.org/abs/1511.07289} {arXiv:1511.07289
  [cs.LG]} \BibitemShut {NoStop}%
\end{thebibliography}%

\end{document}